\documentstyle[graphicx,12pt]{article}
\begin{document}

\small
\hoffset=-1truecm
\voffset=-2truecm
\title{\bf The Casimir force on a piston at finite temperature in Randall-Sundrum models}
\author{Hongbo Cheng\footnote {E-mail address:
hbcheng@sh163.net}\\
Department of Physics, East China University of Science and
Technology,\\ Shanghai 200237, China\\
The Shanghai Key Laboratory of Astrophysics,\\ Shanghai 200234,
China}

\date{}
\maketitle

\begin{abstract}
The Casimir effect for a three-parallel-plate system at finite
temperature within the frame of five-dimensional Randall-Sundrum
models is studied. In the case of Randall-Sundrum model involving
two branes we find that the Casimir force depends on the plates
distance and temperature after one outer plate has been moved to
the distant place. Further we discover that the sign of the
reduced force is negative as the plate and piston are located very
closely, but the reduced force nature becomes repulsive when the
plates distance is not very tiny and finally the repulsive force
vanishes with extremely large plates separation. The higher
temperature causes the repulsive Casimir force greater. Within the
frame of one-brane scenario the reduced Casimir force between the
piston and one plate left keeps attractive no matter how high the
temperature is. It is interesting that stronger thermal effect
leads to greater attractive Casimir force instead of changing the
force nature.
\end{abstract}
\vspace{4cm} \hspace{1cm} PACS number(s): 11.10.Kk, 03.70.+k


\newpage

\noindent \textbf{I.\hspace{0.4cm}Introduction}

More than 80 years ago, the high-dimensional spacetime theory
suggesting that our observable four-dimensional world is a
subspace of a higher dimensional spacetime that has a long
tradition was started by Kaluza and Klein [1, 2]. The
high-dimensional spacetime models including the dimensionality,
topology and the geometric characteristics of extra dimensions are
necessary. The main motivations for such approaches are to unify
all of the fundamental interactions in nature. The issues with
additional dimensions are also invoked for providing a
breakthrough of cosmological constant and the hierarchy problems
[3-8]. These models of high-dimensional spacetimes have their own
compactification and properties of extra dimensions. More theories
need developing and to be realized within the frame with extra
dimensions. In the Kaluza-Klein theory, one extra dimension in our
universe was introduced to be compactified to unify gravity and
classical electrodynamics. The quantum gravity such as string
theories or braneworld scenario is developed to reconcile the
quantum mechanics and gravity with the help of introducing seven
extra spatial dimensions. The new approaches propose that the
strong curvature of the extra spatial dimensions be responsible
for the hierarchy problem. At first, the large extra dimensions
(LED) were put forward [6]. In this model the additional
dimensions are flat and of equal size and the radius of a toroid
is limited to overcome the large gap between the scales of gravity
and electroweak interaction while the size of extra space can not
be too small, or the hierarchy problem remains. Another model with
warped extra dimensions was introduced [7, 8]. A five-dimensional
theory compactified on a $S^{1}/Z_{2}$ manifold, named
Randall-Sundrum (RS) models, suggested that the compact extra
dimension with large curvatures to explain the reason why the
large gap between the Planck and the electroweak scales exists.
Here we choose the RS model as a five-dimensional theory
compactified on a $S^{1}/Z_{2}$ manifold with bulk and boundary
cosmological constants leading to a stable four-dimensional
low-energy effective theory. In RSI, one of the RS models, there
are two 3-branes with equal opposite tensions and they are
localized at $y=0$ and $y=L_{0}$ respectively, with $Z_{2}$
symmetry $y\longleftrightarrow-y$, $L_{0}+y\longleftrightarrow
L_{0}-y$. The Randall-Sundrum model becomes RSII when one brane is
located at infinity, $L_{0}\longrightarrow\infty$. The standard
model field and gauge fields live on the negative tension brane
which is visible, while the positive tension brane with a
fundamental scale $M_{RS}$ is hidden.

The Casimir effect depends on the dimensionality and topology of
the spacetime [9-18] and has received a great deal of attention
within spacetime models including additional spatial dimensions.
There exists strong influence from the size and the geometry of
extra dimensions on the Casimir effect, the evaluation of the
vacuum zero-point energy. The precision of the measurements of the
attractive force between parallel plates as well as other
geometries has been greatly improved practically [19-22], leading
the Casimir effect to be a remarkable observable and trustworthy
consequence of the existence of quantum fluctuations. The
experimental results clearly show that the attractive Casimir
force between the parallel plates vanishes when the plates move
apart from each other to the very distant place. In particular it
must be pointed out that no repulsive force appears. Therefore the
Casimir effect for parallel plates can become a window to probe
the high-dimensional Universe and can be used to research on a
large class of related topics on the various models of spacetime
with more than four dimensions. More efforts have been made on the
studies. Within the frame of several kinds of spacetime with high
dimensionality the Casimir effect for various systems has been
discussed. The eletromagnetic Casimir effect for parallel plates
in a high-dimensional spacetime has been studied and the
subtraction of the divergences in the Casimir energy at the
boundaries is realized [23, 24]. Some topics were studied in the
high-dimensional spacetime described by Kaluza-Klein theory. It is
shown analytically that the extra-dimension corrections to the
Casimir effect for a rectangular cavity in the presence of
compactified universal extra dimensions are very manifest [25].
More attention has been paid to the Casimir effect for the
parallel-plate system in the background governed by Kaluza-Klein
theory [25-37]. It is also proved rigorously that there will
appear repulsive Casimir force between two parallel plates when
the plates distance is sufficiently large in the spacetime with
compactified additional dimensions, and the higher the
dimensionality is, the greater the repulsive force is, unless the
Casimir energy outside the system consisting of two parallel
plates is considered. It should be pointed out that the Casimir
force is modified by the compactified dimensions and the repulsive
part of the modifications has nothing to do with the positions of
the plates, so the repulsive parts of the Casimir force on the
plates must be cancelled. In the case of piston in the same
environment, the Casimir force keeps attractive and more extra
compactified dimensions cause greater attractive force. The
research on the Casimir energy within the frame of Kaluza-Klein
theory to explain the dark energy has been performed and is also
fundamental, and a lot of progresses has been made [38]. In the
context of string theory the Casimir effect was also investigated
[39-42]. Also in the Randall-Sundrum model, the Casimir effect has
been investigated to stabilize the distance between branes
[43-47]. In particular the evaluation of the Casimir force between
two parallel plates under Dirichlet conditions has been performed
in the Randall-Sundrum models with one extra dimension [48-51]. We
declare that the nature of Casimir force between the piston and
its closest plate becomes repulsive in RSI model as the plates
distance is larger enough than the separation between two branes
[51]. In the case of RSII, the Casimir force between piston and
its nearest plate remains attractive while the influence from
warped dimension on the Casimir force between the two parallel
plates is so small that it can be neglected.

The quantum field theory shares many of the effects at finite
temperature. Thermal influence on the Casimir effect is manifest
in many cases [10, 18, 27, 52-57]. The influence from sufficiently
high temperature can even change the conclusions completely. The
stronger thermal influence can lead the Casimir energy to be
positive and the Casimir force to be repulsive in the system
consisting of two parallel plates in the two backgrounds with or
without extra dimensions. The conclusions about Casimir effect for
device with piston in the world such as Randall-Sundrum models
mentioned above are drawn when the temperature is zero. It is
necessary to investigate the Casimir effect for parallel plates in
the Randall-Sundrum models under a nonzero temperature
environment. We must confirm how the thermal influence modifies
the results.

It is fundamental and significant to study the Casimir force on
the piston at finite temperature in the Randall-Sundrum models.
Now we choose a piston device depicted in Fig.1. One plate, called
a piston, is inserted into a two-parallel-plate system and is
parallel to the plates to divide the system into two parts
labelled by A and B respectively. In Part A the distance between
the left plate and the piston is $a$, the remains of the
separation of two original plates is certainly $L-a$, which means
that $L$ denotes the whole plates separation. The total vacuum
energy density of the massless scalar fields obeying Dirichlet
boundary conditions within the region involving a piston shown in
Fig.1 can be written as the sun of three terms,

\begin{equation}
\varepsilon=\varepsilon^{A}(a,T)+\varepsilon^{B}(L-a,T)
+\varepsilon^{out}(T)
\end{equation}

\noindent where $\varepsilon^{A}(a,T)$ and
$\varepsilon^{B}(L-a,T)$ mean the energy density of Part A and B
respectively, and the two terms depend on the temperature and
their own size in these two parts. The term $\varepsilon^{out}(T)$
represents the vacuum energy density outside the system under
thermal influence and is independent of characteristics inside the
system. Having regularized the total vacuum energy density, we
obtain the Casimir energy density,

\begin{equation}
\varepsilon_{C}=\varepsilon_{R}^{A}(a,T)+\varepsilon_{R}^{B}(L-a,T)
+\varepsilon_{R}^{out}(T)
\end{equation}

\noindent where $\varepsilon_{R}^{A}(a,T)$,
$\varepsilon_{R}^{B}(L-a,T)$ and $\varepsilon_{R}^{out}(T)$ denote
the finite parts of terms $\varepsilon^{A}(a,T)$,
$\varepsilon^{B}(L-a,T)$ and $\varepsilon^{out}(T)$ in Eq. (1)
respectively. It should be pointed out that $\varepsilon^{out}(T)$
is not a function of the position of the piston although it
depends on the environment temperature. Further the Casimir force
per unit area on the piston is given with the help of derivative
of the Casimir energy density with respect to the plates distance
like $f'_{C}=-\frac{\partial\varepsilon_{C}}{\partial a}$ and can
be written as,

\begin{equation}
f'_{C}=-\frac{\partial}{\partial a}[\varepsilon_{R}^{A}(a,T)
+\varepsilon_{R}^{B}(L-a,T)]
\end{equation}

\noindent showing that the contribution of vacuum energy from the
exterior region does not modify the Casimir force on the piston.
According to the previous studies, we should point out that the
piston analysis is a correct way to perform the parallel-plate
calculation because we can not neglect the contribution to the
vacuum energy from the area outside the confined region. Further
we wonder how the thermal influence modifies the Casimir effect
for parallel plates in the models. This problem, to our knowledge,
has not been examined. The main purpose of this paper is to
research on the Casimir force between two parallel plates when the
environment temperature does not vanish in the Randall-Sundrum
models. We obtain the Casimir force on a piston in the system
consisting of three parallel plates with nonzero temperature by
means of the zeta-function regularization in the RSI and RSII
models respectively. We also compute the Casimir force in the
limit that one outer plate is moved to the extremely distant
place. We discuss the dependence of the reduced force on the
temperature and compare our results with those with vanishing
temperature and the measurements. Our discussions and conclusions
are listed in the end.

\vspace{3cm}

\noindent \textbf{II.\hspace{0.4cm}The Casimir effect for a piston
at finite temperature in the RSI models}

Here we discuss a massless scalar field living in the bulk at
nonzero temperature in the RS models. Within the frame the
spacetime metric is chosen as,

\begin{equation}
ds^{2}=e^{-2k|y|}g_{\mu\nu}dx^{\mu}dx^{\nu}-dy^{2}
\end{equation}

\noindent where $k$ is assumed to be of the order of the Planck
scale which governs the degree of curvature of the $AdS_{5}$ with
constant negative curvature. That the extra dimension is
compactified on an orbifold gives rise to the generation of the
absolute value of $y$ in the metric. The imaginary time formalism
can be used to describe the scalar fields in the thermal
equilibrium [27, 51-54]. In the five-dimensional RS models we
introduce a partition function for a system,

\begin{equation}
Z=N\int_{periodic}D\Phi\exp[\int_{0}^{\beta}d\tau\int
d^{3}xdy\mathcal{L}(\Phi, \partial_{E}\Phi)]
\end{equation}

\noindent where $\mathcal{L}$ is the Lagrangian density for the
system under consideration, $N$ a constant and "periodic" means
$\Phi(0,x^{\mu},y)=\Phi(\beta,x^{\mu},y)$. Here
$\beta=\frac{1}{T}$ is the inverse of the temperature. In the
five-dimensional spacetime with the background metric denoted in
Eq. (4), the equation of motion for a massless bulk scalar field
$\Phi$ is,

\begin{equation}
g^{\mu\nu}\partial_{\mu}\partial_{\nu}\Phi-e^{2ky}\partial_{y}
(e^{-4ky}\partial_{y}\Phi)=0
\end{equation}

\noindent where $g^{\mu\nu}$ is the usual four-dimensional flat
metric with signature $-2$. The field confining between the two
parallel plates satisfies the Dirichlet boundary conditions
$\Phi(x^{\mu},y)|_{\partial\Omega}=0$, $\partial\Omega$ positions
of the plates in coordinates $x^{\mu}$. Following Ref. [50], we
can choose the $y$-dependent part of the field $\Phi(x^{\mu},y)$
as $\chi^{(N)}(y)$ in virtue of separation of variables.

The general expression for the nonzero modes can be obtained in
terms of the Bessel functions of the first and second kind as,

\begin{equation}
\chi^{(N\neq0)}(y)=e^{2ky}(a_{1}J_{2}(\frac{m_{N}e^{ky}}{k})+a_{2}
Y_{2}(\frac{m_{N}e^{ky}}{k}))
\end{equation}

\noindent where $a_{1}$ and $a_{2}$ are the arbitrary constants.
The effective mass term for the scalar field $m_{N}$ can be
obtained by means of integrating out the fifth dimension $y$. In
the case of RSI model, the hidden and visible 3-branes are located
at $y=0$ and $y=\pi R$ respectively. Here we choose $L_{0}=\pi R$.
According to the modified Neumann boundary conditions
$\frac{\partial\chi^{(N)}}{\partial
y}|_{y=0}=\frac{\partial\chi^{(N)}}{\partial y}|_{y=\pi R}=0$, a
general reduced equation reads,

\begin{equation}
m=m_{N}\approx\kappa(N+\frac{1}{4})
\end{equation}

\noindent where

\begin{equation}
\kappa=\pi ke^{-\pi kR}
\end{equation}

\noindent here we assume $N\gg1$ or equivalently $\pi kR\gg1$
throughout our work.

The modes of the vacuum for parallel plates under the Dirichlet
and modified Neumann boundary conditions for plate positions and
brane locations respectively as mentioned above in RSI at finite
temperature can be expressed as,

\begin{equation}
\omega_{nNl}=\sqrt{p^{2}+(\frac{n\pi}{D})^{2}+m_{N}^{2}+(\frac{2\pi
l}{\beta})^{2}}
\end{equation}

\noindent where

\begin{equation}
p^{2}=p_{1}^{2}+p_{2}^{2}
\end{equation}

\noindent where $p_{1}$ and $p_{2}$ are the wave vectors in the
directions of the unbound space coordinates parallel to the plates
surface and $d$ is the distance of the plates. Here $n$ and $N$
represent positive integers and $l$ stands for an integer. The
generalized zeta function reads,

\begin{eqnarray}
\zeta_{I}(s;-\partial_{E})=Tr(-\partial_{E})^{-s}\hspace{6cm}\nonumber\\
=\int
d^{2}k\sum_{n=1}^{\infty}\sum_{N=1}^{\infty}\sum_{l=-\infty}^{\infty}
[k^{2}+\frac{n^{2}\pi^{2}}{D^{2}}+\kappa^{2}(N+\frac{1}{4})^{2}
+(\frac{2l\pi}{\beta})^{2}]^{-s}
\end{eqnarray}

\noindent where
$\partial_{E}=\frac{\partial^{2}}{\partial\tau^{2}}+\nabla^{2}$
with $\tau=it$. Furthermore, Eq.(12) can also be expressed in
terms of the zeta functions of Epstein-Hurwitz type,

\begin{eqnarray}
\zeta_{I}(s;-\partial_{E})\hspace{10.5cm}\nonumber\\
=\frac{2\pi\Gamma(s-1)}{\Gamma(s)}E_{3}(s-1;\frac{\pi^{2}}{D^{2}},
\kappa^{2},\frac{4\pi^{2}}{\beta^{2}};0,\frac{1}{4},0)\hspace{5.5cm}\nonumber\\
-\frac{2\pi\Gamma(s-1)}{\Gamma(s)}E_{2}^{\frac{\kappa^{2}}{16}}(s-1;
\frac{\pi^{2}}{D^{2}},\frac{4\pi^{2}}{\beta^{2}};0,0)
-\frac{\pi\Gamma(s-1)}{\Gamma(s)}E_{2}(s-1;\frac{\pi^{2}}{D^{2}},\kappa^{2};
0,\frac{1}{4})\nonumber\\
+\frac{\pi\Gamma(s-1)}{\Gamma(s)}E_{1}^{\frac{\kappa^{2}}{16}}(s-1;\frac{\pi^{2}}{D^{2}};0)
-\frac{2\pi\Gamma(s-1)}{\Gamma(s)}E_{2}(s-1;\kappa^{2},\frac{4\pi^{2}}{\beta^{2}};
\frac{1}{4},0)\hspace{1cm}\nonumber\\
+\frac{2\pi\Gamma(s-1)}{\Gamma(s)}E_{1}^{\frac{\kappa^{2}}{16}}(s-1;
\frac{4\pi^{2}}{\beta^{2}};0)+\frac{\pi\Gamma(s-1)}{\Gamma(s)}\kappa^{2-2s}
\zeta_{H}(s-1,\frac{1}{4})\hspace{1.5cm}
\end{eqnarray}

\noindent where the zeta functions of Epstein-Hurwitz type are
defined as,

\begin{equation}
E_{p}^{b_{1}b_{2}\cdot\cdot\cdot
b_{p}}(s;a_{1},a_{2},\cdot\cdot\cdot,a_{p};c_{1},c_{2},\cdot\cdot\cdot,
c_{p})=\sum_{\{n_{j}\}=0}^{\infty}\{\sum_{j=1}^{p}[a_{j}(n_{j}+c_{j})^{2}
+b_{j}]^{-s}\}
\end{equation}

\begin{equation}
E_{p}(s;a_{1},a_{2},\cdot\cdot\cdot,a_{p};c_{1},c_{2},\cdot\cdot\cdot,c_{p})
=\sum_{\{n_{j}\}=0}^{\infty}(\sum_{j=1}^{p}a_{j}(n_{j}+c_{j})^{2})^{-s}
\end{equation}

\noindent and $\zeta_{H}(s,q)=\sum_{n=0}^{\infty}(n+q)^{-s}$ is
the Hurwitz zeta function. The energy density of the
two-parallel-plate system with thermal corrections is,

\begin{equation}
\varepsilon_{I}(D,T)=-\frac{1}{2}\frac{\partial}{\partial\beta}(\frac{\partial\zeta_{I}(s;-\partial_{E})}
{\partial s}|_{s=0})
\end{equation}

\noindent We regularize the expression of vacuum energy density of
the system containing parallel plates to obtain its finite part at
nonzero temperature in RSI model as follows,

\begin{eqnarray}
\varepsilon_{IR}(D,T)\hspace{13cm}\nonumber\\
=-\frac{\Gamma(4)}{64\pi^{\frac{5}{2}}\Gamma(\frac{5}{2})}
\kappa^{3}\sum_{n=1}^{\infty}\frac{\cos n\pi}{(2n)^{4}}
-\frac{\pi^{\frac{3}{2}}}{512\Gamma(\frac{5}{2})}\kappa^{3}\hspace{7.5cm}\nonumber\\
-\frac{\kappa^{2}}{2D}\sum_{n_{1}=1}^{\infty}\sum_{n_{2}=0}^{\infty}
n_{1}^{-2}(n_{2}+\frac{1}{4})^{2}K_{2}(2n_{1}\kappa
D(n_{2}+\frac{1}{4}))
+\frac{\pi}{32}\frac{\kappa^{2}}{D}\sum_{n=1}^{\infty}n^{-2}K_{2}
(\frac{\kappa D}{2\sqrt{\pi}}n)\hspace{1.5cm}\nonumber\\
+2^{\frac{1}{2}}\pi^{\frac{1}{2}}\beta^{-\frac{3}{2}}
\sum_{n_{1}=1}^{\infty}\sum_{n_{2}=0}^{\infty}\sum_{n_{3}=0}^{\infty}
n_{1}^{-\frac{3}{2}}[\frac{n_{2}^{2}\pi^{2}}{D^{2}}+\kappa^{2}(n_{3}+\frac{1}{4})^{2}]^{\frac{3}{4}}
K_{\frac{3}{2}}[n_{1}\beta\sqrt{\frac{\pi^{2}n_{2}^{2}}{D^{2}}+\kappa^{2}(n_{3}+\frac{1}{4})^{2}}]
\nonumber\\-2^{\frac{1}{2}}\pi^{2}\beta^{-\frac{3}{2}}
\sum_{n_{1},n_{2}=1}^{\infty}(\frac{\frac{\pi^{2}n_{1}^{2}}{D^{2}}+\frac{\kappa^{2}}{16}}{\pi^{2}n_{1}^{2}})^{\frac{3}{4}}
K_{\frac{3}{2}}(n_{1}\beta\sqrt{\frac{\pi^{2}n_{2}^{2}}{D^{2}}+\frac{\kappa^{2}}{16}})\hspace{4.5cm}\nonumber\\
+2^{\frac{1}{2}}\pi^{\frac{1}{2}}\beta^{-\frac{1}{2}}
\sum_{n_{1}=1}^{\infty}\sum_{n_{2},n_{3}=0}^{\infty}n_{1}^{-\frac{1}{2}}
[\frac{\pi^{2}n_{2}^{2}}{D^{2}}+\kappa^{2}(n_{3}+\frac{1}{4})^{2}]^{\frac{5}{4}}
[K_{\frac{1}{2}}(n_{1}\beta\sqrt{\frac{\pi^{2}n_{2}^{2}}{D^{2}}+\kappa^{2}(n_{3}+\frac{1}{4})^{2}})\nonumber\\
+K_{\frac{5}{2}}(n_{1}\beta\sqrt{\frac{\pi^{2}n_{2}^{2}}{D^{2}}+\kappa^{2}(n_{3}+\frac{1}{4})^{2}})]
\hspace{6cm}\nonumber\\
-2^{\frac{1}{2}}\pi^{\frac{1}{2}}\beta^{-\frac{1}{2}}
\sum_{n_{1},n_{2}=1}^{\infty}n_{1}^{-\frac{1}{2}}(\frac{\pi^{2}n_{2}^{2}}{D^{2}}
+\frac{\kappa^{2}}{16})^{\frac{5}{4}}[K_{\frac{1}{2}}(n_{1}\beta\sqrt{\frac{\pi^{2}n_{2}^{2}}{D^{2}}+\frac{\kappa^{2}}{16}})
\hspace{3.5cm}\nonumber\\+K_{\frac{5}{2}}(n_{1}\beta\sqrt{\frac{\pi^{2}n_{2}^{2}}{D^{2}}+\frac{\kappa^{2}}{16}})]
\hspace{7cm}\nonumber\\
-8\pi^{-\frac{1}{2}}\Gamma(\frac{3}{2})\zeta(3)\beta^{-3}
\sum_{n=1}^{\infty}\frac{\sin\frac{n\pi}{2}}{n}-12\pi\beta^{-4}\Gamma(2)\zeta(4)
\hspace{5cm}
\end{eqnarray}

\noindent where $K_{\nu}(z)$ is the modified Bessel function of
the second kind. We replace the variable $D$ in Eq. (17) with $a$
and $L-a$ to obtain the Casimir energy densities of Part A and
Part B like,

\begin{eqnarray}
\varepsilon_{IR}^{A}(a,T)=\varepsilon_{IR}(a, T)\hspace{1.5cm}\nonumber\\
\varepsilon_{IR}^{B}(L-a,T)=\varepsilon_{IR}(L-a,T)
\end{eqnarray}

\noindent We obtain the Casimir force per unit area on the piston
at finite temperature in the cosmological background like RSI
model as follows,

\begin{eqnarray}
f'_{IC}=-\frac{\partial}{\partial
a}[\varepsilon^{A}_{IR}(a,T)+\varepsilon^{B}_{IR}(L-a,T)]\hspace{6.5cm}\nonumber\\
=-\frac{\kappa^{2}}{2a^{2}}\sum_{n_{1}=1}^{\infty}\sum_{n_{2}=0}^{\infty}
n_{1}^{-2}(n_{2}+\frac{1}{4})^{2}K_{2}[2n_{1}\kappa a(n_{2}+\frac{1}{4})]\hspace{5cm}\nonumber\\
-\frac{\kappa^{3}}{2a}\sum_{n_{1}=1}^{\infty}\sum_{n_{2}=0}^{\infty}
n_{1}^{-1}(n_{2}+\frac{1}{4})^{3}[K_{1}(2n_{1}\kappa
a(n_{2}+\frac{1}{4})
+K_{3}(2n_{1}\kappa a(n_{2}+\frac{1}{4})))]\hspace{1cm}\nonumber\\
+\frac{1}{32}\frac{\kappa^{3}}{a}\sum_{n=1}^{\infty}
n^{-2}K_{2}(\frac{\kappa
a}{2\sqrt{\pi}}n)+\frac{\pi^{\frac{1}{2}}}{128}
\frac{\kappa^{3}}{a}\sum_{n=1}^{\infty}n^{-1}[K_{1}(\frac{\kappa
a}{2\sqrt{\pi}}n)
+K_{3}(\frac{\kappa a}{2\sqrt{\pi}}n)]\hspace{1cm}\nonumber\\
+\frac{3\sqrt{2}\pi^{\frac{5}{2}}}{2}\frac{1}{a^{\frac{5}{2}}\beta^{\frac{3}{2}}}
\sum_{n_{1}=1}^{\infty}\sum_{n_{2},n_{3}=0}^{\infty}n_{1}^{-\frac{3}{2}}n_{2}^{2}
[\pi^{2}n_{2}^{2}+\kappa^{2}a^{2}(n_{3}+\frac{1}{4})^{2}]^{-\frac{1}{4}}\hspace{2.5cm}\nonumber\\\times
K_{\frac{3}{2}}[n_{1}\frac{\beta}{a}\sqrt{\pi^{2}n_{2}^{2}+\kappa^{2}a^{2}
(n_{3}+\frac{1}{4})^{2}}]\hspace{5cm}\nonumber\\
+2\sqrt{2}\pi^{\frac{5}{2}}\frac{1}{a^{\frac{7}{2}}\beta^{\frac{1}{2}}}
\sum_{n_{1}=1}^{\infty}\sum_{n_{2},n_{3}=0}^{\infty}n_{1}^{-\frac{1}{2}}
n_{2}^{2}[\pi^{2}n_{2}^{2}+\kappa^{2}a^{2}(n_{3}+\frac{1}{4})^{2}]^{\frac{1}{4}}\hspace{3cm}\nonumber\\
\times(K_{\frac{1}{2}}[n_{1}\frac{\beta}{a}\sqrt{\pi^{2}n_{2}^{2}+\kappa^{2}a^{2}(n_{3}+\frac{1}{4})^{2}}]
+K_{\frac{5}{2}}[n_{1}\frac{\beta}{a}\sqrt{\pi^{2}n_{2}^{2}+\kappa^{2}a^{2}(n_{3}+\frac{1}{4})^{2}}])\nonumber\\
-\frac{3\sqrt{2}\pi^{\frac{5}{2}}}{2}\frac{1}{a^{\frac{5}{2}}\beta^{\frac{3}{2}}}
\sum_{n_{1},n_{2}=1}^{\infty}n_{1}^{-\frac{3}{2}}n_{2}^{2}
(\pi^{2}n_{2}^{2}+\frac{\kappa^{2}a^{2}}{16})^{-\frac{1}{4}}
K_{\frac{3}{2}}(n_{1}\frac{\beta}{a}\sqrt{\pi^{2}n_{2}^{2}+\frac{\kappa^{2}a^{2}}{16}})\hspace{1cm}\nonumber\\
-2\sqrt{2}\pi^{\frac{5}{2}}\frac{1}{a^{\frac{7}{2}}\beta^{\frac{1}{2}}}
\sum_{n_{1},n_{2}=1}^{\infty}n_{1}^{-\frac{1}{2}}n_{2}^{2}
(\pi^{2}n_{2}^{2}+\frac{\kappa^{2}a^{2}}{16})^{\frac{1}{4}}\hspace{5cm}\nonumber\\
\times[K_{\frac{1}{2}}(n_{1}\frac{\beta}{a}\sqrt{\pi^{2}n_{2}^{2}+\frac{\kappa^{2}a^{2}}{16}})
+K_{\frac{5}{2}}(n_{1}\frac{\beta}{a}\sqrt{\pi^{2}n_{2}^{2}+\frac{\kappa^{2}a^{2}}{16}})]\hspace{2cm}\nonumber\\
-\frac{\sqrt{2}\pi^{\frac{5}{2}}}{2}\frac{\beta^{\frac{1}{2}}}{\kappa^{\frac{1}{2}}a^{5}}
\sum_{n_{1}=1}^{\infty}\sum_{n_{2},n_{3}=0}^{\infty}n_{1}^{\frac{1}{2}}n_{2}^{2}
[\pi^{2}n_{2}^{2}+\kappa^{2}a^{2}(n_{3}+\frac{1}{4})^{2}]^{\frac{3}{4}}\hspace{3.5cm}\nonumber\\
\times[K_{-\frac{1}{2}}(n_{1}\frac{\beta}{a}\sqrt{\pi^{2}n_{2}^{2}+\kappa^{2}a^{2}(n_{3}+\frac{1}{4})^{2}})
+2K_{\frac{3}{2}}(n_{1}\frac{\beta}{a}\sqrt{\pi^{2}n_{2}^{2}+\kappa^{2}a^{2}(n_{3}+\frac{1}{4})^{2}})\nonumber\\
+K_{\frac{7}{2}}(n_{1}\frac{\beta}{a}\sqrt{\pi^{2}n_{2}^{2}+\kappa^{2}a^{2}(n_{3}+\frac{1}{4})^{2}})]\hspace{4cm}\nonumber\\
+\frac{\sqrt{2}\pi^{\frac{5}{2}}}{2}\frac{\beta^{\frac{1}{2}}}{\kappa^{\frac{1}{2}}a^{5}}
\sum_{n_{1},n_{2}=1}^{\infty}n_{1}^{\frac{1}{2}}n_{2}^{2}
(\pi^{2}n_{2}^{2}+\frac{\kappa^{2}a^{2}}{16})^{\frac{3}{4}}\hspace{6cm}\nonumber\\
\times[K_{-\frac{1}{2}}(n_{1}\frac{\beta}{a}\sqrt{\pi^{2}n_{2}^{2}+\frac{\kappa^{2}a^{2}}{16}})
+2K_{\frac{3}{2}}(n_{1}\frac{\beta}{a}\sqrt{\pi^{2}n_{2}^{2}+\frac{\kappa^{2}a^{2}}{16}})\hspace{2cm}\nonumber\\
+K_{\frac{7}{2}}(n_{1}\frac{\beta}{a}\sqrt{\pi^{2}n_{2}^{2}+\frac{\kappa^{2}a^{2}}{16}})]\hspace{6cm}\nonumber\\
+\frac{\kappa^{2}}{2(L-a)^{2}}\sum_{n_{1}=1}^{\infty}\sum_{n_{2}=0}^{\infty}
n_{1}^{-2}(n_{2}+\frac{1}{4})^{2}K_{2}[2n_{1}\kappa(L-a)(n_{2}+\frac{1}{4})]\hspace{3cm}\nonumber\\
+\frac{\kappa^{3}}{2(L-a)}\sum_{n_{1}=1}^{\infty}\sum_{n_{2}=0}^{\infty}
n_{1}^{-1}(n_{2}+\frac{1}{4})^{3}[K_{1}(2n_{1}\kappa(L-a)(n_{2}+\frac{1}{4})\hspace{3cm}\nonumber\\
+K_{3}(2n_{1}\kappa(L-a)(n_{2}+\frac{1}{4})))]\hspace{4cm}\nonumber\\
-\frac{1}{32}\frac{\kappa^{3}}{L-a}\sum_{n=1}^{\infty}
n^{-2}K_{2}(\frac{\kappa(L-a)}{2\sqrt{\pi}}n)\hspace{7cm}\nonumber\\
-\frac{\pi^{\frac{1}{2}}}{128}
\frac{\kappa^{3}}{L-a}\sum_{n=1}^{\infty}n^{-1}[K_{1}(\frac{\kappa(L-a)}{2\sqrt{\pi}}n)
+K_{3}(\frac{\kappa(L-a)}{2\sqrt{\pi}}n)]\hspace{3.5cm}\nonumber\\
-\frac{3\sqrt{2}\pi^{\frac{5}{2}}}{2}\frac{1}{(L-a)^{\frac{5}{2}}\beta^{\frac{3}{2}}}
\sum_{n_{1}=1}^{\infty}\sum_{n_{2},n_{3}=0}^{\infty}n_{1}^{-\frac{3}{2}}n_{2}^{2}
[\pi^{2}n_{2}^{2}+\kappa^{2}(L-a)^{2}(n_{3}+\frac{1}{4})^{2}]^{-\frac{1}{4}}\hspace{0.5cm}\nonumber\\\times
K_{\frac{3}{2}}[n_{1}\frac{\beta}{L-a}\sqrt{\pi^{2}n_{2}^{2}+\kappa^{2}(L-a)^{2}
(n_{3}+\frac{1}{4})^{2}}]\hspace{3cm}\nonumber\\
-2\sqrt{2}\pi^{\frac{5}{2}}\frac{1}{(L-a)^{\frac{7}{2}}\beta^{\frac{1}{2}}}
\sum_{n_{1}=1}^{\infty}\sum_{n_{2},n_{3}=0}^{\infty}n_{1}^{-\frac{1}{2}}
n_{2}^{2}[\pi^{2}n_{2}^{2}+\kappa^{2}(L-a)^{2}(n_{3}+\frac{1}{4})^{2}]^{\frac{1}{4}}\hspace{1cm}\nonumber\\
\times(K_{\frac{1}{2}}[n_{1}\frac{\beta}{L-a}\sqrt{\pi^{2}n_{2}^{2}+\kappa^{2}(L-a)^{2}(n_{3}+\frac{1}{4})^{2}}]\hspace{3cm}\nonumber\\
+K_{\frac{5}{2}}[n_{1}\frac{\beta}{L-a}\sqrt{\pi^{2}n_{2}^{2}+\kappa^{2}(L-a)^{2}(n_{3}+\frac{1}{4})^{2}}])\hspace{2cm}\nonumber\\
+\frac{3\sqrt{2}\pi^{\frac{5}{2}}}{2}\frac{1}{(L-a)^{\frac{5}{2}}\beta^{\frac{3}{2}}}
\sum_{n_{1},n_{2}=1}^{\infty}n_{1}^{-\frac{3}{2}}n_{2}^{2}
(\pi^{2}n_{2}^{2}+\frac{\kappa^{2}(L-a)^{2}}{16})^{-\frac{1}{4}}\hspace{3cm}\nonumber\\
\times K_{\frac{3}{2}}(n_{1}\frac{\beta}{L-a}\sqrt{\pi^{2}n_{2}^{2}+\frac{\kappa^{2}(L-a)^{2}}{16}})\hspace{4cm}\nonumber\\
+2\sqrt{2}\pi^{\frac{5}{2}}\frac{1}{(L-a)^{\frac{7}{2}}\beta^{\frac{1}{2}}}
\sum_{n_{1},n_{2}=1}^{\infty}n_{1}^{-\frac{1}{2}}n_{2}^{2}
(\pi^{2}n_{2}^{2}+\frac{\kappa^{2}(L-a)^{2}}{16})^{\frac{1}{4}}\hspace{3cm}\nonumber\\
\times[K_{\frac{1}{2}}(n_{1}\frac{\beta}{L-a}\sqrt{\pi^{2}n_{2}^{2}+\frac{\kappa^{2}(L-a)^{2}}{16}})\hspace{5cm}\nonumber\\
+K_{\frac{5}{2}}(n_{1}\frac{\beta}{L-a}\sqrt{\pi^{2}n_{2}^{2}+\frac{\kappa^{2}(L-a)^{2}}{16}})]\hspace{4cm}\nonumber\\
+\frac{\pi^{\frac{5}{2}}}{\sqrt{2}}\frac{\beta^{\frac{1}{2}}}{\kappa^{\frac{1}{2}}(L-a)^{5}}
\sum_{n_{1}=1}^{\infty}\sum_{n_{2},n_{3}=0}^{\infty}n_{1}^{\frac{1}{2}}n_{2}^{2}
[\pi^{2}n_{2}^{2}+\kappa^{2}(L-a)^{2}(n_{3}+\frac{1}{4})^{2}]^{\frac{3}{4}}\hspace{1.5cm}\nonumber\\
\times[K_{-\frac{1}{2}}(n_{1}\frac{\beta}{L-a}\sqrt{\pi^{2}n_{2}^{2}+\kappa^{2}(L-a)^{2}(n_{3}+\frac{1}{4})^{2}})\hspace{3cm}\nonumber\\
+2K_{\frac{3}{2}}(n_{1}\frac{\beta}{L-a}\sqrt{\pi^{2}n_{2}^{2}+\kappa^{2}(L-a)^{2}(n_{3}+\frac{1}{4})^{2}})\hspace{2cm}\nonumber\\
+K_{\frac{7}{2}}(n_{1}\frac{\beta}{L-a}\sqrt{\pi^{2}n_{2}^{2}+\kappa^{2}(L-a)^{2}(n_{3}+\frac{1}{4})^{2}})]\hspace{2cm}\nonumber\\
-\frac{\pi^{\frac{5}{2}}}{\sqrt{2}}\frac{\beta^{\frac{1}{2}}}{\kappa^{\frac{1}{2}}(L-a)^{5}}
\sum_{n_{1},n_{2}=1}^{\infty}n_{1}^{\frac{1}{2}}n_{2}^{2}
(\pi^{2}n_{2}^{2}+\frac{\kappa^{2}(L-a)^{2}}{16})^{\frac{3}{4}}\hspace{3.5cm}\nonumber\\
\times[K_{-\frac{1}{2}}(n_{1}\frac{\beta}{L-a}\sqrt{\pi^{2}n_{2}^{2}+\frac{\kappa^{2}(L-a)^{2}}{16}})\hspace{3cm}\nonumber\\
+2K_{\frac{3}{2}}(n_{1}\frac{\beta}{L-a}\sqrt{\pi^{2}n_{2}^{2}+\frac{\kappa^{2}(L-a)^{2}}{16}})\hspace{3cm}\nonumber\\
+K_{\frac{7}{2}}(n_{1}\frac{\beta}{L-a}\sqrt{\pi^{2}n_{2}^{2}+\frac{\kappa^{2}(L-a)^{2}}{16}})]\hspace{3cm}
\end{eqnarray}

\noindent This expression represents the Casimir pressure on the
piston before the right plate of the system depicted in Fig. 1 has
been moved to the remote place. Further we take the limit
$L\longrightarrow\infty$ which means that the right plate in Part
B is moved to a very distant place, then we obtain the following
expression for the Casimir force per unit area on the piston
within the frame of two-brane Randall-Sundrum issue,

\begin{eqnarray}
f_{IC}=\lim_{L\longrightarrow\infty}f'_{IC}\hspace{9.5cm}\nonumber\\
=-\frac{\kappa^{4}}{2\mu^{2}}\sum_{n_{1}=1}^{\infty}\sum_{n_{2}=0}^{\infty}
n_{1}^{-2}(n_{2}+\frac{1}{4})^{2}K_{2}[2n_{1}\mu(n_{2}+\frac{1}{4})]\hspace{4cm}\nonumber\\
-\frac{\kappa^{4}}{2\mu}\sum_{n_{1}=1}^{\infty}\sum_{n_{2}=0}^{\infty}
n_{1}^{-1}(n_{2}+\frac{1}{4})^{3}[K_{1}(2n_{1}\mu(n_{2}+\frac{1}{4})
+K_{3}(2n_{1}\mu(n_{2}+\frac{1}{4})))]\nonumber\\
+\frac{\pi}{32}\frac{\kappa^{4}}{\mu^{2}}\sum_{n=1}^{\infty}
n^{-2}K_{2}(\frac{\mu}{2\sqrt{\pi}}n)+\frac{\pi^{\frac{1}{2}}}{128}
\frac{\kappa^{4}}{\mu}\sum_{n=1}^{\infty}n^{-1}[K_{1}(\frac{\mu}{2\sqrt{\pi}}n)
+K_{3}(\frac{\mu}{2\sqrt{\pi}}n)]\nonumber\\
+\frac{3\sqrt{2}\pi^{\frac{5}{2}}}{2}\frac{\kappa^{4}}{\mu^{\frac{5}{2}}\xi^{\frac{3}{2}}}
\sum_{n_{1}=1}^{\infty}\sum_{n_{2},n_{3}=0}^{\infty}n_{1}^{-\frac{3}{2}}n_{2}^{2}
[\pi^{2}n_{2}^{2}+\mu^{2}(n_{3}+\frac{1}{4})^{2}]^{-\frac{1}{4}}\hspace{2cm}\nonumber\\
\times
K_{\frac{3}{2}}[n_{1}\frac{\xi}{\mu}\sqrt{\pi^{2}n_{2}^{2}+\mu^{2}
(n_{3}+\frac{1}{4})^{2}}]\hspace{3cm}\nonumber\\
+2\sqrt{2}\pi^{\frac{5}{2}}\frac{\kappa^{4}}{\mu^{\frac{7}{2}}\xi^{\frac{1}{2}}}
\sum_{n_{1}=1}^{\infty}\sum_{n_{2},n_{3}=0}^{\infty}n_{1}^{-\frac{1}{2}}
n_{2}^{2}[\pi^{2}n_{2}^{2}+\mu^{2}(n_{3}+\frac{1}{4})^{2}]^{\frac{1}{4}}\hspace{2.5cm}\nonumber\\
\times(K_{\frac{1}{2}}[n_{1}\frac{\xi}{\mu}\sqrt{\pi^{2}n_{2}^{2}+\mu^{2}(n_{3}+\frac{1}{4})^{2}}]
+K_{\frac{5}{2}}[n_{1}\frac{\xi}{\mu}\sqrt{\pi^{2}n_{2}^{2}+\mu^{2}(n_{3}+\frac{1}{4})^{2}}])\nonumber\\
-\frac{3\sqrt{2}\pi^{\frac{5}{2}}}{2}\frac{\kappa^{4}}{\mu^{\frac{5}{2}}\xi^{\frac{3}{2}}}
\sum_{n_{1},n_{2}=1}^{\infty}n_{1}^{-\frac{3}{2}}n_{2}^{2}(\pi^{2}n_{2}^{2}+\frac{\mu^{2}}{16})^{-\frac{1}{4}}
K_{\frac{3}{2}}(n_{1}\frac{\xi}{\mu}\sqrt{\pi^{2}n_{2}^{2}+\frac{\mu^{2}}{16}})\hspace{0.5cm}\nonumber\\
-2\sqrt{2}\pi^{\frac{5}{2}}\frac{\kappa^{4}}{\mu^{\frac{7}{2}}\xi^{\frac{1}{2}}}
\sum_{n_{1},n_{2}=1}^{\infty}n_{1}^{-\frac{1}{2}}n_{2}^{2}(\pi^{2}n_{2}^{2}+\frac{\mu^{2}}{16})^{\frac{1}{4}}\hspace{4.5cm}\nonumber\\
\times[K_{\frac{1}{2}}(n_{1}\frac{\xi}{\mu}\sqrt{\pi^{2}n_{2}^{2}+\frac{\mu^{2}}{16}})
+K_{\frac{5}{2}}(n_{1}\frac{\xi}{\mu}\sqrt{\pi^{2}n_{2}^{2}+\frac{\mu^{2}}{16}})]\hspace{1cm}\nonumber\\
-\frac{\pi^{\frac{5}{2}}}{\sqrt{2}}\frac{\kappa^{4}\xi^{\frac{1}{2}}}{\mu^{5}}
\sum_{n_{1}=1}^{\infty}\sum_{n_{2},n_{3}=0}^{\infty}n_{1}^{\frac{1}{2}}n_{2}^{2}
[\pi^{2}n_{2}^{2}+\mu^{2}(n_{3}+\frac{1}{4})^{2}]^{\frac{3}{4}}\hspace{3cm}\nonumber\\
\times[K_{-\frac{1}{2}}(n_{1}\frac{\xi}{\mu}\sqrt{\pi^{2}n_{2}^{2}+\mu^{2}(n_{3}+\frac{1}{4})^{2}})
+2K_{\frac{3}{2}}(n_{1}\frac{\xi}{\mu}\sqrt{\pi^{2}n_{2}^{2}+\mu^{2}(n_{3}+\frac{1}{4})^{2}})\nonumber\\
+K_{\frac{7}{2}}(n_{1}\frac{\xi}{\mu}\sqrt{\pi^{2}n_{2}^{2}+\mu^{2}(n_{3}+\frac{1}{4})^{2}})]\hspace{3cm}\nonumber\\
+\frac{\pi^{\frac{5}{2}}}{\sqrt{2}}\frac{\kappa^{4}\xi^{\frac{1}{2}}}{\mu^{5}}
\sum_{n_{1},n_{2}=1}^{\infty}n_{1}^{\frac{1}{2}}n_{2}^{2}
(\pi^{2}n_{2}^{2}+\frac{\mu^{2}}{16})^{\frac{3}{4}}\hspace{5cm}\nonumber\\
\times[K_{-\frac{1}{2}}(n_{1}\frac{\xi}{\mu}\sqrt{\pi^{2}n_{2}^{2}+\frac{\mu^{2}}{16}})
+2K_{\frac{3}{2}}(n_{1}\frac{\xi}{\mu}\sqrt{\pi^{2}n_{2}^{2}+\frac{\mu^{2}}{16}})\hspace{1cm}\nonumber\\
+K_{\frac{7}{2}}(n_{1}\frac{\xi}{\mu}\sqrt{\pi^{2}n_{2}^{2}+\frac{\mu^{2}}{16}})]\hspace{3cm}
\end{eqnarray}

\noindent while we introduce two dimensionless variables, the
scaled temperature and the relation between plates separation and
the distance between two 3-branes respectively,

\begin{equation}
\xi=\kappa\beta=\pi k\beta e^{-\pi kR}
\end{equation}

\begin{equation}
\mu=\kappa a=\pi kae^{-\pi kR}
\end{equation}

\noindent The terms with series in Eq.(20) converge very quickly
and only the first several summands need to be taken into account
for numerical calculation in further discussions. If the
temperature approaches zero, the Casimir force will recover to be
the results of Ref. [51]. We have to perform the burden and
surprisingly difficult calculation on Eq. (20) in order to explore
the Casimir force on the piston at finite temperature in the
cosmological background governed by the RSI model. It is clear
that the force expression depends on the plate-piston distance and
temperature. For a definite temperature like $\xi=1$, the
numerical evaluations of the Casimir force per unit area on the
piston from Eq. (20) lead to the data presented in Fig. 1. We find
that the sign of the Casimir force is negative when the
dimensionless variable $\mu$ defined in (22) is very tiny. When
the distance between the plate and piston is larger enough than
the branes separation $R$, meaning the value of $\mu$ is
sufficiently large, the nature of the Casimir force turns to be
repulsive although the force vanishes as the plates separation
approaches to infinity like
$\lim_{\mu\longrightarrow\infty}f_{IC}=0$. The curves of the
dependence of the Casimir force per unit area for the piston on
the plates distance for different temperatures are similar. They
possess several general characters such as the attractive Casimir
force with very small $\mu$ or the repulsive one with sufficiently
large $\mu$ and the asymptotic behaviour
$f_{IC}(\mu\longrightarrow\infty,T)=0$. All of the expressions for
the Casimir forced with thermal corrections have positive maxima.
The dependence of the top values of the curves on the scaled
temperature $\xi$ defined in Eq. (21) is shown in Fig. 3. The
higher temperature or equivalently lower scaled values leads to
larger positive top magnitude, which means that the Casimir force
between the plate and piston is an increasing function of
temperature. The thermal influence has not cancelled the positive
nature of Casimir force but results in the stronger repulsive
force. In a word, there also appears the repulsive Casimir force
between two parallel plates inevitably under thermal influence in
the RSI model, which is excluded by the experimental evidence. It
should be emphasized that there appears a term like
$\frac{\pi}{32}\frac{\kappa^{2}}{D}\sum_{n=1}^{\infty}n^{-2}K_{2}
(\frac{\kappa D}{2\sqrt{\pi}}n)$ in our expression (17) and it is
the term that finally leads the Casimir force between two parallel
plates in the RSI model to become repulsive when the plates
separation is not extremely tiny. We also find that our results
involving the term
$\frac{\pi}{32}\frac{\kappa^{2}}{D}\sum_{n=1}^{\infty}n^{-2}K_{2}
(\frac{\kappa D}{2\sqrt{\pi}}n)$ are subject to $m_{N=0}=0$. The
term
$\frac{\pi}{32}\frac{\kappa^{2}}{D}\sum_{n=1}^{\infty}n^{-2}K_{2}
(\frac{\kappa D}{2\sqrt{\pi}}n)$ will not appear if
$m_{N=0}=\frac{\kappa}{4}$ is chosen, then the Casimir energy and
Casimir force will be the same as Frank et al's [50, 53], and
$m_{N=0}=\frac{\kappa}{4}$ is not acceptable here. It should be
pointed out that the equation is valid asymptotically for $N\gg1$
although the reduced equation (8) for the effective mass of the
scalar bulk field is expressed as an approximation. The error is
about $3\%$ when $N=1$ and the error is $0.3\%$ and $0.1\%$ for
$N=2$ and $N=3$ respectively, etc., displaying that the error
drops very quickly with increasing $N$ [53]. The deviation from
the approximation in the case of small $N$ can not change the
above conclusion.

\vspace{3cm}

\noindent \textbf{III.\hspace{0.4cm}The Casimir force for a piston
at finite temperature in the RSII models}

In this section, we proceed with the same study on the Casimir
effect in the RSII model, in which the 3-brane at $y=\pi R$ is at
infinity. That the 3-brane is moved to the infinity leads the
spectrum of the Kaluza-Klein masses to be continuous and run all
$m>0$. The generalized zeta function becomes,

\begin{eqnarray}
\zeta_{II}(s;\partial_{E})=Tr(-\partial_{E})^{-s}\hspace{5cm}\nonumber\\
=\sum_{n=1}^{\infty}\sum_{l=-\infty}^{\infty}\int_{0}^{\infty}
\frac{dm}{k}\int
d^{2}p[p^{2}+(\frac{n\pi}{D})^{2}+m^{2}+(\frac{2l\pi}{\beta})^{2}]^{-s}
\end{eqnarray}

\noindent here the parameter $k$ is the same as in metric (4) and
is determined by the 5D Planck mass and bulk cosmological
constant. Similarly after the integration the generalized zeta
function for RSII model can be expressed with the help of the
Epstein zeta functions as,

\begin{equation}
\zeta_{II}(s;\partial_{E})=\frac{\pi^{\frac{3}{2}}}{k}
\frac{\Gamma(s-\frac{3}{2})}{\Gamma(s)}E_{2}(s-\frac{3}{2};\frac{\pi^{2}}{D^{2}},\frac{4\pi^{2}}{\beta^{2}})
+\frac{\pi^{\frac{3}{2}}}{2k}\frac{\Gamma(s-\frac{3}{2})}{\Gamma(s)}
(\frac{\pi}{D})^{3-2s}\zeta(2s-3)
\end{equation}

\noindent where

\begin{equation}
E_{p}(s;a_{1},a_{2},\cdot\cdot\cdot,a_{p})=\sum_{\{n_{j}\}=1}^{\infty}
(\sum_{j=1}^{p}a_{j}n_{j}^{2})^{-s}
\end{equation}

\noindent and $\zeta(s)$ is the Riemann zeta function. Similarly
the vacuum energy density of device involving two parallel plates
at finite temperature in the RSII scenario is,

\begin{equation}
\varepsilon_{II}(D,T)=-\frac{1}{2}\frac{\partial}{\partial\beta}(\frac{\partial\zeta_{II}(s;-\partial_{E})}
{\partial s}|_{s=0})
\end{equation}

\noindent We regularize the expressions to obtain the finite parts
of the vacuum energy density for parallel plates in the RSII model
when the world temperature does not vanish,

\begin{eqnarray}
\varepsilon_{IIR}(D,T)=-\frac{\sqrt{\pi}}{8}\frac{1}{kD^{4}}
\Gamma(\frac{5}{2})\zeta(5)+\frac{2\pi^{3}}{k\beta^{2}D^{2}}
\sum_{n_{1},n_{2}=1}^{\infty}(\frac{n_{2}}{n_{1}})^{2}
K_{2}(\frac{\pi\beta}{D}n_{1}n_{2})\nonumber\\
+\frac{\pi^{4}}{k\beta D^{3}}\sum_{n_{1},n_{2}=1}^{\infty}
\frac{n_{2}^{3}}{n_{1}}[K_{1}(\frac{\pi\beta}{D}n_{1}n_{2})
+K_{3}(\frac{\pi\beta}{D}n_{1}n_{2})]\hspace{1cm}\nonumber\\
\end{eqnarray}

\noindent Now we choose the variable $D$ in Eq. (27) as $a$ and
$L-a$ respectively to obtain the Casimir energy densities of Part
A and Part B as follows,

\begin{eqnarray}
\varepsilon_{IIR}^{A}(a,T)=\varepsilon_{IIR}(a,T)\hspace{1.5cm}\nonumber\\
\varepsilon_{IIR}^{B}(L-a,T)=\varepsilon_{IIR}(L-a,T)
\end{eqnarray}

\noindent According to Eq. (3), the Casimir per unit area on the
piston belonging to a three-parallel-plate system in the RSII
model introduces,

\begin{eqnarray}
f'_{IIC}=-\frac{\partial}{\partial a}[\varepsilon_{IIR}^{A}(a,T)
+\varepsilon_{IIR}^{B}(L-a,T)]\hspace{5.5cm}\nonumber\\
=-\frac{\sqrt{\pi}}{2ka^{5}}\Gamma(\frac{5}{2})\zeta(5)
+\frac{4\pi^{3}}{k\beta^{2}a^{3}}\sum_{n_{1},n_{2}=1}^{\infty}
(\frac{n_{2}}{n_{1}})^{2}K_{2}(\frac{\pi\beta}{a}n_{1}n_{2})\hspace{4cm}\nonumber\\
+\frac{2\pi^{4}}{k\beta a^{4}}\sum_{n_{1},n_{2}=1}^{\infty}
\frac{n_{2}^{3}}{n_{1}}[K_{1}(\frac{\pi\beta}{a}n_{1}n_{2})
+K_{3}(\frac{\pi\beta}{a}n_{1}n_{2})]\hspace{4.5cm}\nonumber\\
-\frac{\pi^{5}}{2ka^{5}}\sum_{n_{1},n_{2}=1}^{\infty}
[K_{0}(\frac{\pi\beta}{a}n_{1}n_{2})+2K_{2}(\frac{\pi\beta}{a}n_{1}n_{2})
+K_{4}(\frac{\pi\beta}{a}n_{1}n_{2})]\hspace{2cm}\nonumber\\
+\frac{\sqrt{\pi}}{2k(L-a)^{5}}\Gamma(\frac{5}{2})\zeta(5)
-\frac{8\pi^{3}}{k\beta^{2}(L-a)^{3}}\sum_{n_{1},n_{2}=1}^{\infty}
(\frac{n_{2}}{n_{1}})^{2}K_{2}(\frac{\pi\beta}{L-a}n_{1}n_{2})\hspace{1cm}\nonumber\\
-\frac{2\pi^{4}}{k\beta (L-a)^{4}}\sum_{n_{1},n_{2}=1}^{\infty}
\frac{n_{2}^{3}}{n_{1}}[K_{1}(\frac{\pi\beta}{L-a}n_{1}n_{2})
+K_{3}(\frac{\pi\beta}{a}n_{1}n_{2})]\hspace{3cm}\nonumber\\
+\frac{\pi^{5}}{2k(L-a)^{5}}\sum_{n_{1},n_{2}=1}^{\infty}
[K_{0}(\frac{\pi\beta}{L-a}n_{1}n_{2})+2K_{2}(\frac{\pi\beta}{L-a}n_{1}n_{2})
+K_{4}(\frac{\pi\beta}{L-a}n_{1}n_{2})]
\end{eqnarray}

\noindent In order to show the Casimir force between the piston
and its closer plate and compare our conclusions with the
measurements, we let $L\longrightarrow\infty$ to find,

\begin{eqnarray}
f_{IIC}=-\frac{\sqrt{\pi}}{2ka^{5}}\Gamma(\frac{5}{2})\zeta(5)
+\frac{4\pi^{3}}{k\beta^{2}a^{3}}\sum_{n_{1},n_{2}=1}^{\infty}
(\frac{n_{2}}{n_{1}})^{2}K_{2}(\frac{\pi\beta}{a}n_{1}n_{2})\hspace{4cm}\nonumber\\
+\frac{2\pi^{4}}{k\beta a^{4}}\sum_{n_{1},n_{2}=1}^{\infty}
\frac{n_{2}^{3}}{n_{1}}[K_{1}(\frac{\pi\beta}{a}n_{1}n_{2})
+K_{3}(\frac{\pi\beta}{a}n_{1}n_{2})]\hspace{4cm}\nonumber\\
-\frac{\pi^{5}}{2ka^{5}}\sum_{n_{1},n_{2}=1}^{\infty}
[K_{0}(\frac{\pi\beta}{a}n_{1}n_{2})+2K_{2}(\frac{\pi\beta}{a}n_{1}n_{2})
+K_{4}(\frac{\pi\beta}{a}n_{1}n_{2})]\hspace{2cm}
\end{eqnarray}

\noindent The dependence of the reduced Casimir force per unit
area on the plate-piston separation with some values of
temperature is plotted in Fig. 4. If the thermal influence is
omitted, the above expression of the reduced Casimir pressure will
be recovered to be the findings in Ref. [51], just containing a
deviation from the results of the conventional parallel-plate
system. As the temperature is high enough, i.e.
$\beta\longrightarrow0$, then

\begin{equation}
f_{IIC}(\beta\longrightarrow0)=-\frac{16\sqrt{\pi}}{k\beta^{5}}
\Gamma(\frac{5}{2})\zeta(5)
\end{equation}

\noindent It is clear that the magnitude of Casimir force on the
piston increases with the fifth power of temperature. It also
indicates that the sign of the reduced force keeps negative, which
means that the plate and piston still attract each other, while
higher temperature certainly gives rise to greater attractive
Casimir force instead of causing the reduced force to be
repulsive. In the four-dimensional flat spacetime the sign of the
Casimir force will change to be positive when the temperature is
large enough. Our results about the nature of Casimir force
between the piston and the remain plate at finite temperature in
the context of RSII model are different from those in the
background whose dimensionality is four, which are not disfavoured
by the measurements.

\vspace{3cm}

\noindent \textbf{IV.\hspace{0.4cm}Conclusions}

The Casimir force between two parallel plates involving the
contribution from exterior vacuum energy with thermal corrections
is studied in the presence of one warped extra dimension of the
models proposed by Randall and Sundrum. In the two-brane scenario
called RSI model we derive the Casimir force at finite temperature
for the three-parallel-plate system where the middle plate is
called piston. We get the exact form of reduced Casimir force per
unit area between one plate and the piston as one outer plate is
moved away. In this limiting case we find that the sign of the
reduced force depending on the temperature and distance between
the plate and the piston will become positive when the
plate-piston gap is not extremely tiny although the force will
disappear as the piston and plate leave far from each other. The
stronger thermal influence brings on greater repulsive Casimir
force between the plates. In the case of RSI model at finite
temperature a repulsive Casimir force is produced due to warps
between the parallel plates and the repulsive force is associated
with the plates distance, so the repulsive parts of the Casimir
force on the piston can not be cancelled although the repulsive
parts will vanish when the two parallel plates move away. The
appearance of the repulsive Casimir force between one plate and
the piston conflicts with the experimental results. It is obvious
that the RSI model cannot be reliable according to our analysis
even we consider the thermal influence during our research.

In the case of one brane called RSII model we perform the same
study and procedure to find the reduced Casimir force per unit
area on the piston. We find that the reduced force with thermal
corrections is great when the piston and plate are located very
closely each other or vanishes with very large plate-piston
distance while the force always keeps attractive, no matter how
high the temperature is. It is interesting that stronger thermal
influence gives rises to greater attractive Casimir force, instead
of changing the force to be repulsive.

\vspace{3cm}

\noindent\textbf{Acknowledgement}

This work was supported by NSFC No. 10875043 and partly supported
by the Shanghai Research Foundation No. 07dz22020.

\newpage
\begin{figure}
\setlength{\belowcaptionskip}{10pt} \centering
  \includegraphics[width=15cm]{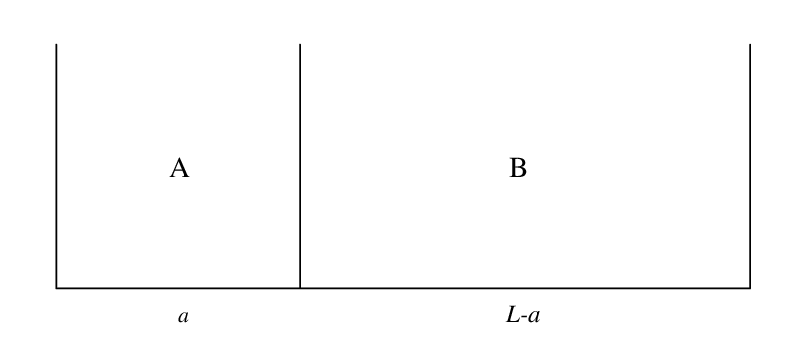}
  \caption{Casimir piston}
\end{figure}

\newpage
\begin{figure}
\setlength{\belowcaptionskip}{10pt} \centering
  \includegraphics[width=15cm]{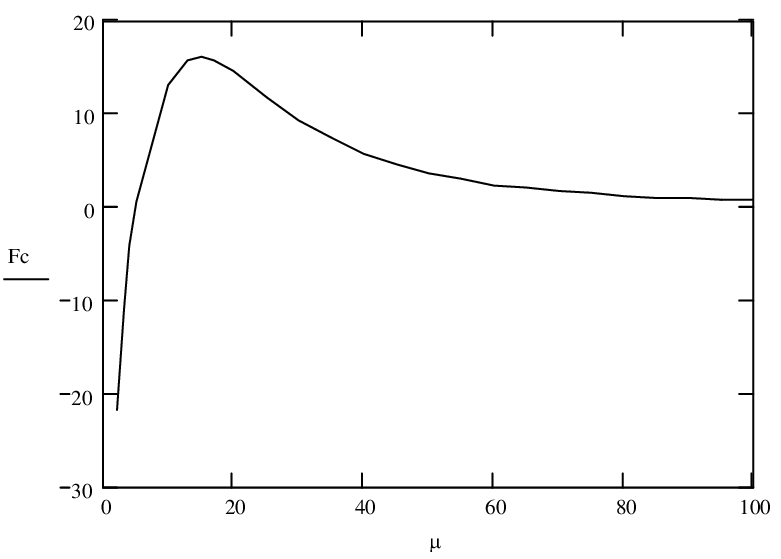}
  \caption{The Casimir force per unit area in unit of $\kappa^{4}$
  between the plate and piston versus the dimensionless variable denoted as $\mu=\kappa a$ when $\xi=1$}
\end{figure}

\newpage
\begin{figure}
\setlength{\belowcaptionskip}{10pt} \centering
  \includegraphics[width=15cm]{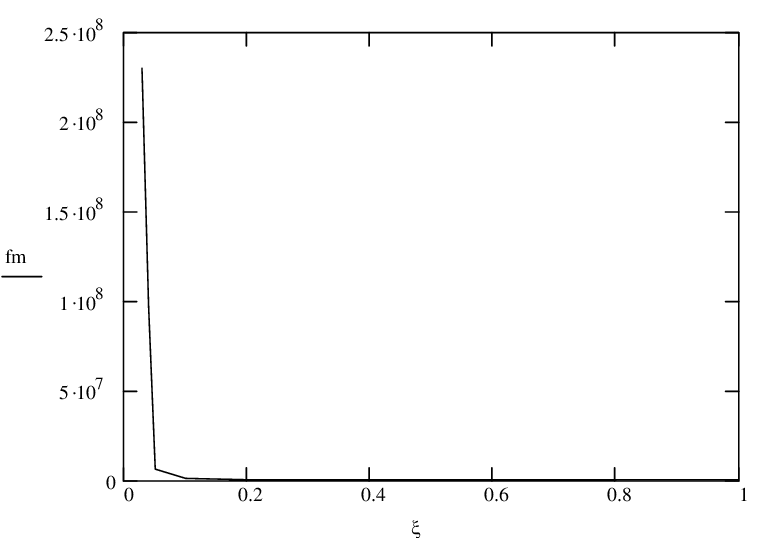}
  \caption{The dependence of the top values of the Casimir force per unit area in unit of $\kappa^{4}$
  between the plate and piston on the scaled temperature $\xi=\kappa\beta$}
\end{figure}

\newpage
\begin{figure}
\setlength{\belowcaptionskip}{10pt} \centering
  \includegraphics[width=15cm]{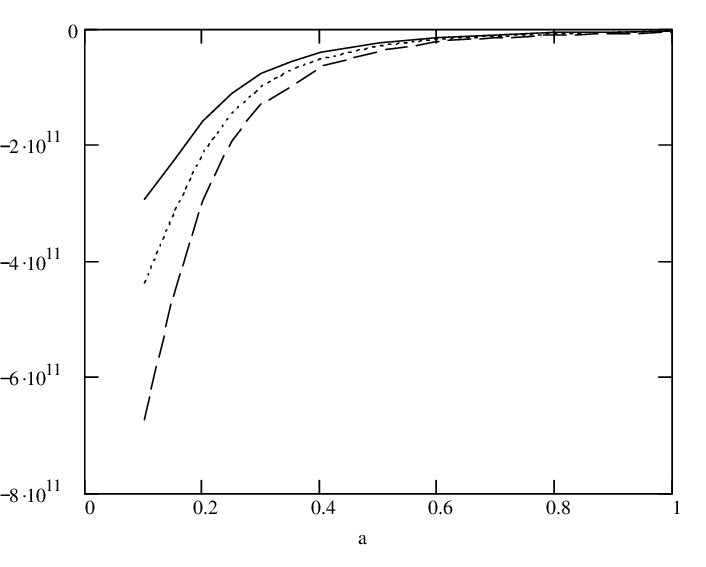}
  \caption{The dashed, dot and solid curves of Casimir force per unit area on the piston as functions
  of plate-piston distance in 5-dimensional RSII model for $\beta=0.01,0.011,0.012$ respectively.}
\end{figure}

\end{document}